\documentclass{article}
\usepackage{amsmath,amssymb,graphicx}
\usepackage{cite}
\newcommand{\J}{\mathrm{J}}
\newcommand{\st}{\mathrm{st}}
\newcommand{\dy}{\mathrm{dy}}
\newcommand{\rd}{\mathrm{d}}
\begin{document}
\begin{titlepage}
\begin{center}
{\large \textbf{The spectrum and the phase transition of models solvable through the full interval method }}

\vspace{2\baselineskip}
{\sffamily Amir~Aghamohammadi~\footnote{e-mail: mohamadi@alzahra.ac.ir},
Mohammad~Khorrami~\footnote{e-mail: mamwad@mailaps.org}.}

\vspace{2\baselineskip}
{\it Department of Physics, Alzahra University, Tehran 19938-93973, Iran}
\end{center}
\vspace{2\baselineskip}
\begin{abstract}
\noindent The most general exclusion single species
reaction-diffusion models with nearest-neighbor interactions one a
one dimensional lattice are investigated, for which the evolution of
full intervals are closed. Using a generating function method,
the probability that $n$ consecutive sites be full is investigated.
The stationary values of these probabilities, as well as the spectrum
of the time translation generator (Hamiltonian) of these are obtained.
It is shown that depending on the reaction rates the model could
exhibit a dynamical phase transition.
\end{abstract}
\end{titlepage}
\newpage
\section{Introduction}
Different methods have been used to study  non equilibrium statistical models;
including analytical and asymptotic methods, mean-field methods, and
large-scale numerical methods. Although mean field solutions may be suitable
for higher dimensions, in low dimensional cases fluctuation effects may cause
mean field results to differ from the real ones. One dimensional models,
which are usually easier to investigate, helps us gaining more knowledge
on systems far from equilibrium
\cite{ScR,HHL1,MHP2,ADHR,KPWH,HS1,PCG,HOS1,HOS2,AL,AKK,AKK2,AM1}. Many techniques
are used to obtain exact results in one-dimensional models. 
The empty interval method (EIM) is one of them. The full interval method
(FIM) is another, which is equivalent to EIM. In  EIM, the aim is to calculate 
the probability that $n$ consecutive sites be empty, $E_n$. In FIM, 
one studies the probability that $n$ consecutive sites be full, $F_n$,). 
In \cite{BDb,BDb1,BDb2,BDb3}, one dimensional models have been studied in
which some of the reaction rates are infinite, so that it is the (finite) 
diffusion rates which determine the evolution. These models have been 
through EIM. In \cite{HH}, a system with three-site interactions has 
been studied by EIM. In \cite{Mb}, a generalization of EIM has been
used to investigate a one dimensional Potts model with
$q$-states, in the zero-temperature limit.
In \cite{AKA}, a classification has been given for one dimensional 
reaction-diffusion models with nearest neighbor interactions,
which are solvable through EIM, by which it is meant that 
the evolution equations for $E_n$'s are closed. In such systems, 
the time derivative of the empty intervals is linear in the empty
intervals (and contains no other dynamical variable). If some extra 
conditions are met (there is no reaction which produces 
particles in two adjacent sites) the evolution equation for $E_n$ 
becomes $n$-independent. This makes solving the evolution equations
easier. In \cite{KAA,AAK,AK}, these were generalized to systems 
with multi-species systems and multi-site interactions. In \cite{AK2}, 
the extra condition that the evolution equations for $E_n$ be
$n$-independent was relaxed, and solvable (in the sense of EIM) models 
on continuum were studied. 

In \cite{MMPB}, a generalization of
EIM was introduced, by which a model was investigated which was not
solvable through conventional EIM. In \cite{DFBHR,DFH}, EIM has been used 
to study the coagulation-diffusion process on a one dimensional chain.

In a recent article, the most general exclusion single species
one dimensional reaction-diffusion models with nearest-neighbor
interactions were studied, which are both autonomous and can be
solved exactly through the full interval method \cite{MH}. There,
using a generating function method, the general solution for $F_n$,
the probability that $n$ consecutive sites be full, and some other
correlation functions of number operators were explicitly obtained.

In this article we relax the condition of autonomy, and study
the most general exclusion single species reaction-diffusion models
with nearest-neighbor interactions on a one dimensional lattice,
which can be solved exactly through full interval method.
The change of empty interval to full interval is, of course,
not important, as a simple interchange of particles and
holes would do that. The scheme of the paper is as follows.
In Section 2, the most general exclusion single species
one dimensional reaction-diffusion models with nearest-neighbor
interactions are introduced, which can be solved exactly
through FIM. In Section 3, a generating function method is used
to obtain the time independent solutions to the evolution of
full intervals. In Section 4, a similar generating function
method is used to study the spectrum of the time translation
generator (Hamiltonian) of the full intervals. it is shown that
the model may exhibit a dynamical phase transition.
\section{Full interval equation}
Consider a one dimensional lattice, any site of which is either
occupied by a single particle or empty. The full interval
$F_n$ is defined as the probability that $n$ consecutive sites be full
\begin{equation}\label{04.1}
F_n:=P(\overbrace{\bullet\bullet\cdots\bullet }^n).
\end{equation}
An empty (occupied) site is denoted by by $\circ$ ($\bullet$).
Assuming only nearest neighbor interactions, it can be seen (\cite{AK2})
that the necessary and sufficient condition that the evolution
equations for $F_n$'s be closed is that onl the following reactions
be present, and their rates be related to each other as follows.
\begin{align}\label{04.2}
\circ\bullet&\to\begin{cases}\circ\circ,& q_1\\
                              \bullet\circ,& r_1
\end{cases},\nonumber\\
\bullet\circ&\to\begin{cases}\circ\circ,& q_2\\
                           \circ\bullet,& r_2
\end{cases},\nonumber\\
\circ\circ&\to\begin{cases}\bullet\circ,& r_1\\
                    \circ\bullet,& r_2
\end{cases},\nonumber\\
\bullet\bullet&\to\begin{cases}\bullet\circ,&w_1\\
                              \circ\bullet,&w_2\\
                               \circ\circ,&w
\end{cases}.
\end{align}
The equality of some rates, means that the rates of the reactions
$(\circ\circ\to\bullet\circ)$ and $(\circ\bullet\to\bullet\circ)$
are the same, and the rates of the reactions
$(\circ\circ\to\circ\bullet)$ and $(\bullet\circ\to\circ\bullet)$
are the same as well.

As in, for example \cite{AK2}, it can be seen that the time evolution equation for $F_n$ is
\begin{align}\label{04.3}
\frac{\rd F_n}{\rd t}&=(r_1+r_2)\,(F_{n-1}+F_{n+1}-2\,F_n)-(q_1+q_2)\,(F_n-F_{n+1})\nonumber\\
&\quad-(n-1)\,(w_1+w_2+w)\,F_n-(w_1+w_2+2\,w)\,F_{n+1},\qquad n\geq 2\nonumber\\
\frac{\rd F_1}{\rd t}&=(r_1+r_2)\,(1+F_2-2\,F_1)-(q_1+q_2)\,(F_1-F_2)\nonumber\\
&\quad-(w_1+w_2+2\,w)\,F_2.
\end{align}
It is seen that defining
\begin{equation}\label{04.4}
F_0:=1,
\end{equation}
the evolution equation for $F_1$ takes a form similar to that of
other $F_n$'s:
\begin{align}\label{04.5}
\frac{\rd F_n}{\rd t}&=(r_1+r_2)\,(F_{n-1}+F_{n+1}-2\,F_n)-(q_1+q_2)\,(F_n-F_{n+1})\nonumber\\
&\quad-(n-1)\,(w_1+w_2+w)\,F_n-(w_1+w_2+2\,w)\,F_{n+1},\qquad n\geq 1.
\end{align}
Comparing this with the similar expression in, it is seen that 
in \cite{AK2} there is no term analogues to the last term here. 
In \cite{MH}, it is assumed the models to be autonomous, which leads to
removing the term $F_{n+1}$ from (\ref{04.5}).

Defining
\begin{equation}\label{04.6}
\tilde t:=(w_1+w_2+w)\,t,
\end{equation}
one arrives at
\begin{equation}\label{04.7}
\frac{\rd F_n}{\rd\tilde t}=b\,F_{n-1}-(a+n-1)\,F_n+c\,F_{n+1},
\end{equation}
where
\begin{align}\label{04.8}
a&:=\frac{2\,(r_1+r_2)+q_1+q_2}{w+w_1+w_2},\nonumber\\
b&:=\frac{r_1+r_2}{w+w_1+w_2},\nonumber\\
c&:=\frac{r_1+r_2+q_1+q_2-2\,w-w_1-w_2}{w+w_1+w_2}.
\end{align}
The case $(w_1+w_2+w)=0$, will be dealt with separately.
For simplicity, hereafter the symbol $t$ is used in place of $\tilde t$,
so
\begin{equation}\label{04.9}
\frac{\rd F_n}{\rd t}=b\,F_{n-1}-(a+n-1)\,F_n+c\,F_{n+1}.
\end{equation}
A generating function $F$ is also defined, which will be used later:
\begin{equation}\label{04.10}
F(x):=\sum_{n=0}^\infty\frac{F_n\,x^n}{n!}.
\end{equation}
It is seen that
\begin{equation}\label{04.11}
F_n=F^{(n)}(0),
\end{equation}
where $F^{(n)}$ is the $n$-th derivative of $F$.
\section{The time independent equation}
The time independent solution $F^\st$ satisfies
\begin{equation}\label{04.12}
b\,F^\st_{n-1}-(a+n-1)\,F^\st_n+c\,F^\st_{n+1}=0,
\end{equation}
which yields
\begin{equation}\label{04.13}
\sum_{n=1}^\infty\frac{x^{n-1}}{(n-1)!}\,[c\,F^\st_{n+1}+b\,F^\st_{n-1}-(a+n-1)\,F^\st_n]=0,
\end{equation}
so that
\begin{equation}\label{04.14}
(c-x)\,\frac{\rd^2 F^\st}{\rd x^2}-a\,\frac{\rd F^\st}{\rd x}+b\,F^\st=0.
\end{equation}
Defining
\begin{align}\label{04.15}
F^\st(x)&:=z^{(1-a)}\,G^\st(z),\nonumber\\
z&:=2\,\sqrt{b\,(c-x)},
\end{align}
one arrives at
\begin{equation}\label{04.16}
z^2\,\frac{\rd^2 G^\st}{\rd z^2}+z\,\frac{\rd G^\st}{\rd z}+[z^2-(1-a)^2]\,G^\st=0.
\end{equation}
So,
\begin{equation}\label{04.17}
G^\st(z)=\alpha\,\J_{a-1}(z)+\beta\,\J_{1-a}(z),
\end{equation}
where $\J_\nu$ is the Bessel function or order $\nu$,
and $\alpha$ and $\beta$ are constants. So,
\begin{align}\label{04.18}
F^\st(x)&=\alpha\,[2\sqrt{b\,(c-x)}]^{(1-a)}\,\J_{a-1}[2\sqrt{b\,(c-x)}]\nonumber\\
&\quad+\beta\,[2\sqrt{b\,(c-x)}]^{(1-a)}\,\J_{1-a}[2\sqrt{b\,(c-x)}].
\end{align}
As $F_n^\st$ is in $[0,1]$, the convergence radius of the series
defing $F^\st$ is infinity. So the the generating function $F^\st$
as a function of $x$ is analytic on the entire complex plain. Using
\begin{equation}\label{04.19}
\J_\nu(z)=\frac{1}{\Gamma(\nu+1)}\,\left(\frac{z}{2}\right)^\nu+\cdots,\qquad|z|\ll 1,
\end{equation}
the analyticity of $F^\st$ at $(x=c)$ demands that
\begin{equation}\label{04.20}
\beta=0.
\end{equation}
As
\begin{equation}\label{04.21}
F^\st_0=1,
\end{equation}
one has
\begin{equation}\label{04.22}
F^\st(0)=1,
\end{equation}
which can be exploited to obtain $\alpha$. So,
\begin{equation}\label{04.23}
F^\st(x)=\frac{[2\sqrt{b\,(c-x)}]^{(1-a)}\,\J_{a-1}[2\sqrt{b\,(c-x)}]}
{(2\sqrt{b\,c})^{(1-a)}\,\J_{a-1}(2\sqrt{b\,c})}.
\end{equation}

Using (\ref{04.11}), one can find $F_n$. One has
\begin{align}\label{04.24}
\left(\frac{\rd}{\rd x}\right)^n F^\st(x)&=
\frac{1}{(2\,\sqrt{b\,c})^{1-a}\,\J_{a-1}(2\,\sqrt{b\,c})}\,
\left(-\frac{2\,b}{z}\,\frac{\rd}{\rd z}\right)^n[z^{1-a}\,\J_{a-1}(z)],\nonumber\\
&=\frac{1}{(2\,\sqrt{b\,c})^{1-a}\,\J_{a-1}(2\,\sqrt{b\,c})}\,\,(2\,b)^n\,
z^{1-a-n}\,\J_{a-1+n}(z),
\end{align}
which results in
\begin{equation}\label{04.25}
F^\st_n=\left(\frac{b}{c}\right)^{n/2}\,\frac{\J_{a-1+n}(2\,\sqrt{b\,c})}{\J_{a-1}(2\,\sqrt{b\,c})}.
\end{equation}
There are two limiting cases to be studied separately:\\[\baselineskip]
\textbf{i}: $b=0$.\\
This can be studied as the limit $b\to 0$ of the general case.
Using (\ref{04.25}) and the limiting behavior
(\ref{04.19}), one arrives at
\begin{equation}\label{04.26}
F^\st_n=\frac{\Gamma(a)}{\Gamma(a+n)}\,b^n,\qquad b\ll 1,
\end{equation}
which leads to
\begin{equation}\label{04.27}
\lim_{b\to 0}F^\st_n=\delta^0_n.
\end{equation}
This is the case if $a$ is nonzero. If $a$ and $b$ both
vanish, the result could be obtained directly from (\ref{04.12})
to be
\begin{equation}\label{04.28}
F^\st_n=\begin{cases}
\rho,& n=1\\
0,& n>1
\end{cases},
\end{equation}
where
\begin{equation}\label{04.29}
0\leq\rho\leq\frac{1}{2},
\end{equation}
and it has been assumed that at least one of the rates are
nonvanishing. One notes that $\rho$ is in fact the density of
the particles, and the restriction on its value results from
the fact that in the stationary configuration no two-adjacent sites
are full.

These results are expected. If $b$ vanishes but $a$ does not,
there are no reactions which produce particles but there are
reactions which annihilate particles, whether the particles
are adjacent to holes or other particles. So at large time
the lattice becomes empty. If both $a$ and $b$ vanish,
there are no reactions which produce particles, but there are
reactions which annihilate particles, only if there are
two neighboring particles. So particles will be annihilated,
but when there are particles with empty neighboring sites,
they will survive.

In fact, regarding the stationary solution as the large-time solution,
one can obtain an expression for $\rho$ in terms of the initial
conditions. As both $a$ and $b$ vanish, one has
\begin{equation}\label{40.30}
\frac{\rd F_n}{\rd t}=-(n-1)\,F_n+c\,F_{n+1}.
\end{equation}
Defining
\begin{equation}\label{40.31}
\mathcal{F}_n(t):=\exp[(n-1)\,t]\,F_n(t),
\end{equation}
equation (\ref{40.30}) is recast to
\begin{equation}\label{40.32}
\frac{\rd\mathcal{F}_n}{\rd t}=c\,\exp(-t)\,\mathcal{F}_{n+1},
\end{equation}
which results is
\begin{equation}\label{40.33}
\mathcal{F}_n(t)=\sum_{k=0}^\infty\frac{[c-c\,\exp(-t)]^k}{k!}\,\mathcal{F}_{n+k}(0),
\end{equation}
so that
\begin{equation}\label{40.34}
F_n(t)=\exp[-(n-1)\,t]\,\sum_{k=0}^\infty\frac{[c-c\,\exp(-t)]^k}{k!}\,F_{n+k}(0),
\end{equation}
from which,
\begin{equation}\label{40.35}
F_1(\infty)=\sum_{k=0}^\infty\frac{c^k}{k!}\,F_{1+k}(0),
\end{equation}
or,
\begin{equation}\label{40.36}
\rho=\sum_{k=0}^\infty\frac{c^k}{k!}\,F_{1+k}(0).
\end{equation}
One notes that when $a$ and $b$ both vanish, $c$ is nonpositive.
\\[\baselineskip]
\textbf{ii}: $w_1+w_2+w=0$.\\
In this case all of the rates $w_1$, $w_2$, and $w$ should vanish.
One arrives at
\begin{align}\label{40.37}
&(r_1+r_2)\,F^\st_{n-1}-(2\,r_1+2\,r_2+q_1+q_2)\,F^\st_n+(r_1+r_2+q_1+q_2)\,F^\st_{n+1}=0,\nonumber\\
&F^\st_0=1.
\end{align}
The solution to (\ref{40.37}) is
\begin{equation}\label{40.38}
F^\st_n=\zeta+(1-\zeta)\,\left(\frac{r_1+r_2}{r_1+r_2+q_1+q_2}\right)^n,
\end{equation}
where
\begin{equation}\label{40.39}
0\leq\zeta\leq 1,
\end{equation}
provided of $q_1$ and $q_2$, at least one is nonvanishing. Otherwise
\begin{equation}\label{40.40}
F^\st_n=1,
\end{equation}
which corresponds to a full lattice.
\section{Relaxation towards the time independent equation}
To study the spectrum of the time translation generator
(Hamiltonian) of the full intervals, again the generating function
is used. Defining
\begin{equation}\label{40.41}
F^\dy_n:=F_n-F^\st_n,
\end{equation}
one arrives at
\begin{equation}\label{40.42}
\frac{\rd F^\dy_n}{\rd t}=b\,F^\dy_{n-1}-(a+n-1)\,F^\dy_n+c\,F^\dy_{n+1},
\end{equation}
with the boundary condition
\begin{equation}\label{40.43}
F^\dy_0=0.
\end{equation}
Equation (\ref{40.42}) is of the form
\begin{equation}\label{40.44}
\frac{\rd F^\dy_n}{\rd t}=(h\,F^\dy)_n,
\end{equation}
where $h$ is a linear operator.
To find this relaxation time, one should obtain the eigenvalues of $h$.
The eigenvalue with the largest real part, determines the relaxation time.
Denoting the eigenvector of $h$ corresponding to the eigenvalue $E$ by
$\psi_E$, one has
\begin{equation}\label{40.45}
E\,\psi_{E\,n}=b\,\psi_{E\,n-1}-(a+n-1)\,\psi_{E\,n}+c\,\psi_{E\,n+1},
\end{equation}
where $E$ is the corresponding eigenvalue. This is similar to
(\ref{04.12}), with $a$ replaced by $(a+E)$. So repeating similar
arguments, one arrives at
\begin{equation}\label{40.46}
\psi_E(x)=\alpha\,(c-x)^{(1-a-E)/2}\,\J_{a+E-1}[2\sqrt{b\,(c-x)}].
\end{equation}
But now the boundary condition is
\begin{equation}\label{40.47}
\psi_E(0)=0,
\end{equation}
which results in
\begin{equation}\label{40.48}
\J_{a+E-1}(2\sqrt{b\,c})=0.
\end{equation}
This gives the spectrum of $h$.

In the case $b=0$, one can find more explicit forms for
the eigenvalues and eigenvectors. Starting from
(\ref{40.45}), one arrives at
\begin{equation}\label{40.49}
\psi_{E\,n+1}=\frac{E+a+n-1}{c}\,\psi_{E\,n}.
\end{equation}
This recursive relations shows that $\psi_{E\,n}$ tends to
infinity as $n$ tends to infinity, unless there is a $k$ so that
$\psi_{E\,n}$'s vanish for $n>k$. This happens if $E$ is
equal to one of $E_k$'s, where
\begin{equation}\label{40.50}
E_k=1-a-k,
\end{equation}
and $k$ is a positive integer. Denoting the corresponding eigenvector
by $\psi_k$ instead of $\psi_E$, one would arrive at
\begin{equation}\label{40.51}
\psi_{k\,n}=\frac{(-c)^{k-n}}{\Gamma(k+1-n)}\,\psi_{k\,k}.
\end{equation}
The relaxation time of the system is obtained from the largest
real part of the eigenvalues, which in this case is $(-a)$. So
\begin{equation}\label{40.52}
\tau=\frac{1}{a},\qquad (b\,c)=0.
\end{equation}

Defining
\begin{equation}\label{40.53}
\varepsilon:=E+a-1,
\end{equation}
it is seen from (\ref{40.48}) that $\varepsilon$ depends on only
the product $(b\,c)$. So the expression (\ref{40.50}) holds for
the case $c=0$ as well. The case $c=0$ has already been discussed
in greater detail in \cite{MH}

From (\ref{04.8}) is it seen that $b$
is nonnegative, while $c$ can change sign. These expressions also show that
\begin{equation}\label{40.54}
-1\leq(b\,c),
\end{equation}
but there is no upper limit for $(b\,c)$.

For $(b\,c)$ near zero, one can find the leading correction to
(\ref{40.51}) as follows. Defining
\begin{equation}\label{40.55}
\delta_k:=\varepsilon_k+k,
\end{equation}
equation (\ref{40.48}) becomes
\begin{equation}\label{40.56}
\J_{-k+\delta_k}(2\,\sqrt{bc})=0,
\end{equation}
which is, up to the leading order, equivalent to
\begin{equation}\label{40.57}
1-\frac{(b\,c)^k}{k!\,(k-1)!\,\delta_k}=0,
\end{equation}
showing that
\begin{equation}\label{40.58}
\delta_k=\frac{(b\,c)^k}{k!\,(k-1)!}+\cdots,\qquad|b\,c|\ll k^2.
\end{equation}
So one has,
\begin{equation}\label{40.59}
\tau=\frac{1}{a-b\,c}+\cdots,\qquad |b\,c|\ll 1.
\end{equation}

If $(b\,c)$ is positive, there is an inner product with respect to
which $h$ is Hermitian. So if $(b\,c)$ is positive, the spectrum
of $h$ is real. Increasing $(b\,c)$ from zero, the values of
$\varepsilon_k$'s are also increased. For large values of $(b\,c)$
one can find the values of $\varepsilon_k$'s using various asymptotic
expressions of the Bessel functions. One arrives at
\begin{equation}\label{40.60}
\varepsilon_k=\begin{cases}
2\,\sqrt{b\,c}-{\mathfrak{a}}_k\,(b\,c)^{1/6}+\cdots,& k\ll (b\,c)^{1/3}\\ \\
\displaystyle{\frac{4\,\sqrt{b\,c}}{\pi}+\frac{3}{2}-2\,k+\cdots},& |\varepsilon_k|\ll (b\,c)^{1/2}\\ \\
\displaystyle{-k+\frac{(b\,c)^k}{k!\,(k-1)!}+\cdots},& k\gg (b\,c)^{1/2}
\end{cases},
\end{equation}
where $(-\mathfrak{a}_k)$ is the $k$'th zero of the Airy function. The largest
of $\varepsilon_k$'s determine the relaxation time. One then has
\begin{equation}\label{40.61}
\tau=\frac{1}{a-1+2\,\sqrt{b\,c}}+\cdots,\qquad(b\,c)\gg 1.
\end{equation}
From (\ref{04.8}) is it seen that for nonnegative $c$,
\begin{equation}\label{40.62}
(a-1)\geq2\,\sqrt{b\,c}.
\end{equation}

For a negative $(b\,c)$, however, there are cases where the spectrum of
$h$ is not real. A plot of $(b\,c)$ in terms of (real) $\varepsilon$ for
\begin{equation}\label{40.63}
\mathrm{J}_\varepsilon(2\,\sqrt{b\,c})=0,
\end{equation}
shows that there is one minimum for $(b\,c)$ for each interval
$\varepsilon\in(-2\,n,-2\,n+1)$, where $n$ is a positive integer, figure 1.
\begin{figure}
\begin{center}
\includegraphics[scale=0.6]{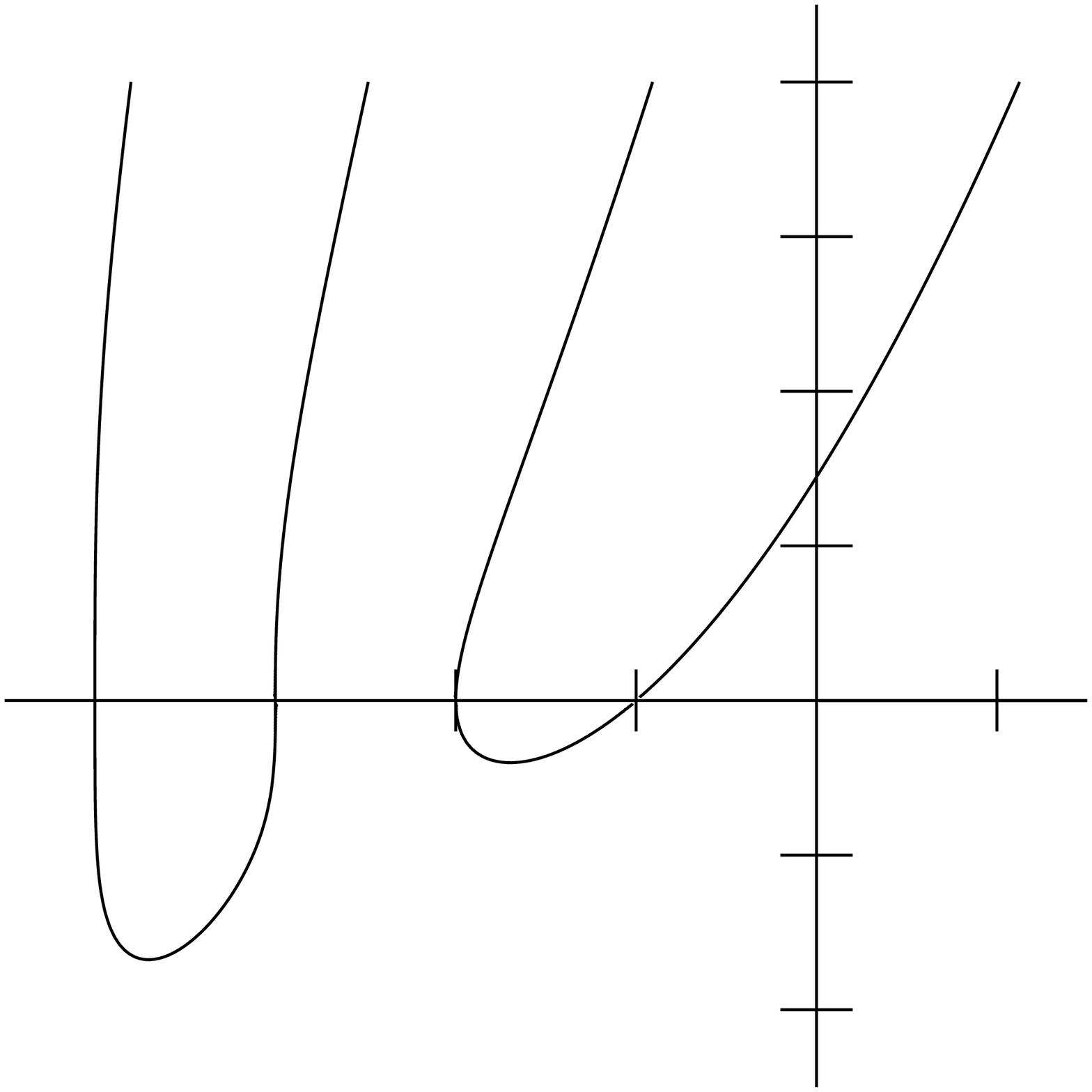}
\setlength{\unitlength}{1pt}
\put(-80,325){\Large$(b\,c)$}
\put(-15,110){\Large$\varepsilon$}
\put(-73,163){\Large 1}
\put(-36,133){\Large 1}
\caption{\label{fig1}}
{The plot of $(b\,c)$ versus $\varepsilon$, corresponding to
the four largest eigenvalues of $h$}
\end{center}
\end{figure}
However, only the minimum corresponding to $\varepsilon\in(-2,-1)$
is larger than $(-1)$. So there is a critical value for $(b\,c)$,
at which two of the eigenvalues of $h$ (the largest and the next largest)
become equal, and if $(b\,c)$ is less than that critical value, two of
the eigenvalues of $h$ become nonreal (complex conjugates of each other.
Denoting that critical value of $(b\,c)$ by $\gamma$, and
the corresponding value of $\varepsilon$ by $\varepsilon_\mathrm{tr}$, one
arrives at the following approximate expression for
$\varepsilon$ and $(b\,c)$ near their critical value.
\begin{equation}\label{40.64}
(b\,c)=\gamma+\frac{(\varepsilon-\varepsilon_\mathrm{c})^2}{\nu},
\end{equation}
where the numerical values of the constants in the above equation are
\begin{align}\label{40.65}
\gamma&=-0.401873,\nonumber\\
\nu&=0.754464,\nonumber\\
\varepsilon_\mathrm{c}&=-1.697524.
\end{align}
One then arrives at the following expression for $\varepsilon_1$, for
$(b\,c)$ near the critical value $\gamma$:
\begin{equation}\label{40.66}
\varepsilon_1=\varepsilon_\mathrm{c}+\sqrt{\nu\,(b\,c-\gamma)},
\end{equation}
resulting to
\begin{equation}\label{40.67}
\tau=\begin{cases}
\displaystyle{\frac{1}{a-1-\varepsilon_\mathrm{c}-\sqrt{\nu\,(b\,c-\gamma)}}},&(b\,c)\gtrsim\gamma\\ \\
\displaystyle{\frac{1}{a-1-\varepsilon_\mathrm{c}}},&(b\,c)\lesssim\gamma
\end{cases}.
\end{equation}
So the derivative of the relaxation time with respect to $(b\,c)$,
is infinite for $(b\,c)\to\gamma^+$, and zero for $(b\,c)\to\gamma^-$.
This model shows a dynamical phase transition.
\\[\baselineskip]
\textbf{Acknowledgement}:  This work was supported by
the research council of the Alzahra University.

\end{document}